# Optimization of Undersampling Parameters for 3D Intracranial Compressed Sensing MR Angiography at 7 Tesla

Matthijs H.S. de Buck[1], Peter Jezzard[1], and Aaron T. Hess[2,3]

[1]Wellcome Centre for Integrative Neuroimaging, FMRIB Division, Nuffield Department of Clinical Neurosciences, University of Oxford, Oxford, United Kingdom, [2]Oxford Centre for Clinical Magnetic Resonance Research, Department of Cardiovascular Medicine, University of Oxford, Oxford, United Kingdom, [3]BHF Centre of Research Excellence, University of Oxford, Oxford, United Kingdom

**Correspondence address**:

Aaron T. Hess, PhD

Oxford Centre for Clinical Magnetic Resonance Research

Department of Cardiovascular Medicine

University of Oxford

John Radcliffe Hospital

Oxford, OX3 9DU

United Kingdom

Email: aaron.hess@cardiov.ox.ac.uk

# Abstract

**Purpose:** 3D Time-of-flight (TOF) MR Angiography (MRA) can accurately visualize the intracranial vasculature, but is limited by long acquisition times. Compressed sensing (CS) reconstruction can be used to substantially accelerate acquisitions. The quality of those reconstructions depends on the undersampling patterns used in the acquisitions. In this work, optimized sets of undersampling parameters using various acceleration factors for Cartesian 3D TOF-MRA are established.

**Methods:** Fully-sampled datasets acquired at 7T were retrospectively undersampled using variable-density Poisson-disk sampling with various autocalibration region sizes, polynomial orders, and acceleration factors. The accuracy of reconstructions from the different undersampled datasets was assessed using the vessel-masked structural similarity index. Results were compared for four imaging volumes, acquired from two different subjects. Optimized undersampling parameters were validated using additional prospectively undersampled datasets.

**Results:** For all acceleration factors, using a fully-sampled calibration area of 12×12 k-space lines and a polynomial order of around 2-2.4 resulted in the highest image quality. The importance of sampling parameter optimization was found to increase for higher acceleration factors. The results were consistent across resolutions and regions of interest with vessels of varying sizes and tortuosity. In prospectively undersampled acquisitions, using optimized undersampling parameters resulted in a 7.2% increase in the number of visible small vessels at R = 7.2.

**Conclusion:** The image quality of CS TOF-MRA can be improved by appropriate choice of undersampling parameters. The optimized sets of parameters are independent of the acceleration factor.

# 1. Introduction

Time-of-flight (TOF) magnetic resonance angiography (MRA) is a valuable technique for clinical study of the intracranial vasculature. It visualizes the blood in a certain region (slice or slab) by generating bright-blood contrast between inflowing and stationary spins. In the brain it can be used for detection of various types of vascular complications, such as atherosclerosis and stenosis [1] or aneurysms [2-3]. Compared to other angiography techniques, such as computed tomography angiography, TOF-MRA has the benefits of being a non-invasive technique, without the need for intravenous contrast agents, and without exposing subjects to ionising radiation.

High spatial resolution MRA can visualize small and highly tortuous vessels such as the lenticulostriate arteries (LSAs) [4], which are implicated in up to a third of symptomatic strokes [5]. The spatial resolution and contrast in MRA can be improved by using ultra-high field MRI, with static magnetic field strengths of ≥7 tesla [6]. TOF-MRA at 7T benefits from longer T1-relaxation times and increased signal-to-noise ratio (SNR), resulting in the potential for higher resolution acquisitions and improved visibility of small vessels [4]. However, the achieved spatial resolution is limited by long acquisition times, which can lead to reduced patient comfort, increased patient movement, and increased clinical costs. In order to remain within clinical scan durations, sub-Nyquist sampling techniques are required, for example using parallel imaging techniques [7].

Compressed sensing (CS) [8] techniques have the potential to achieve high acceleration factors. CS combines highly undersampled non-uniform acquisitions and sparsity in a given domain to restore image quality. Due to the intrinsic sparsity of TOF-MRA data in both the image and wavelet domain, and the improved MRA contrast at 7T, it has already been shown that acceleration factors of 7.2 [9] or higher [10-11] can be achieved with a minimal reduction in clinical image quality [12], or even with improved diagnostic image quality compared to conventional acceleration methods [11].

Cartesian undersampled k-space trajectories for three-dimensional TOF-MRA with CS reconstruction are commonly designed using two-dimensional undersampling covering the two phase-encode directions ($k_y$, $k_z$), with each sampled point in the ($k_y$, $k_z$)-plane representing a continuously sampled line in the frequency-encode direction ($k_x$). Such undersampling masks are often created using pseudo-random variable-density Poisson disks [13] with a fully-sampled calibration region in the centre of the ($k_y$, $k_z$)-plane [9], [12-16]. Variable-density Poisson disk-undersampling distributions are characterized by three parameters: (1) the undersampling factor (*R*), (2) the polynomial order of the sampling density variation (*pp*), and (3) the size of the fully-sampled calibration region (*calib*).

Although the image quality depends on those undersampling parameters, no conclusive information is available about their optimal values for 3D TOF-MRA at 7T. It also remains unclear how the optimal acquisition parameters depend on the acceleration factor and resolution being used. Previous work has compared the image quality from retrospectively undersampled 3D MRA-data for various accelerations factors [10], [14] or for the combination of acceleration factor and calibration region size in undersampled 2D MRI acquisitions for different contrasts [17]. For dynamic MRI [18] and numerical T1-weighted brain models [19], studies into the optimization of compressed sensing and acquisition parameters are available. However, it is unclear how this translates to the case of 3D MRA, which requires the visibility of smaller structures in the reconstructed images, and which has the potential for higher acceleration factors due to the higher intrinsic sparsity.

In this work [20], 3D TOF-MRA undersampling parameters were optimised by retrospectively evaluating different calibration region sizes and polynomial orders at six acceleration factors (from 5 to 15). Finally, the identified undersampling patterns were evaluated in healthy subjects using additional prospectively undersampled acquisitions that adopted the optimized parameters.

# 2. Methods

## 2.1 Data acquisition

Data were acquired using a 3D gradient echo non-contrast-enhanced TOF-MRA sequence on a Siemens (Erlangen, Germany) Magnetom 7T scanner using a 1Tx32Rx head-coil. All experiments were performed under an agreed technical development protocol approved by the Oxford University Clinical Trials and Research Governance office, in accordance with International Electrotechnical Commission and UK Health Protection Agency guidelines. In each subject, 4 sequential slabs [21] consisting of 640×506×56 voxels with a resolution of (0.31 mm)$^3$ were acquired using a distance factor of -19.64% and a combined field-of-view of 200×157×60mm. Further sequence parameters were: TR/TE = 14/5.61 ms (allowing for asymmetric echo), flip angle = 20°, bandwidth = 118 Hz/pixel. To reduce SAR, an increased excitation pulse duration of 1.536 ms was used. The acquisition time was 26:39 minutes for a fully-sampled acquisition.

Fully-sampled datasets from two subjects were used for parameter optimization through retrospective undersampling ("Cohort 1"). After an optimized set of acquisition parameters was established, additional datasets were acquired using a modified version of the TOF-MRA sequence (that incorporated undersampling). A fully sampled acquisition and four different undersampled datasets, using both the optimized and more conventional undersampling parameters, were acquired from six healthy volunteers ("Cohort 2"). The reconstructed image quality for those additional datasets was compared to assess consistency of the findings.

## 2.2 Undersampling

Undersampling masks were generated using the SPIRiT Toolbox v0.3[1], characterised by the three previously described parameters ($R, pp,$ and $calib$). A polynomial order of 0 corresponds to homogeneous undersampling, with higher values corresponding to denser sampling closer to the centre of k-space. Larger calibration regions (consisting of $calib \times calib$ continuously sampled frequency-encode lines) improve the estimation of coil sensitivities [22], but take up more scan time (reducing the effective acceleration). Undersampling was performed retrospectively for the fully-sampled data acquired in Cohort 1 at acceleration factors ranging from 5 to 15. The calibration region size was chosen to range from the lower limit for the ESPIRiT tool (*calib* = 10) to the highest value found in literature for 3D MRA [9]. In Cohort 2, prospectively undersampled data were acquired for validation at R = 7.2 (which was previously found to be "a reasonable trade-off between scan time and image quality" [9]) and R = 15.

---

[1] https://people.eecs.berkeley.edu/~mlustig/Software.html. Note that in this work the acceleration factor refers to the *effective* acceleration (i.e., including the time taken to scan the calibration region), which is different from the definition of the "acc" parameter in the SPIRiT-vdPoisMex tool.

### 2.3 Reconstruction

For all reconstructions, coil sensitivities were estimated using ESPIRiT [22] based on the fully-sampled calibration region in the centre of k-space. For the reconstruction of undersampled data, compressed sensing was implemented using a FISTA-based algorithm [23] with l1-regularization in the wavelet domain [24] using the `pics` (**p**arallel **i**maging and **c**ompressed **s**ensing) tool in the BART-toolbox [25-26] (v0.4.02). The l1-regularization seeks to find the solution to:

$$\min_{\mathbf{x}}\{|\mathbf{A}(\mathbf{x}) - \mathbf{b}| + \lambda|\phi(\mathbf{x})|_1\}, \tag{1}$$

where **b** represents the acquired (undersampled) k-space data, **A** is the undersampling operator over the reconstructed k-space **x**, $\lambda$ is a regularization parameter, and $\phi$ denotes the wavelet transform. The first term ($|\mathbf{A}(\mathbf{x}) - \mathbf{b}|$) ensures data consistency, whereas the second term ($\lambda|\phi(\mathbf{x})|_1$) enforces sparsity in the wavelet domain. Appropriate values for the number of iterations ($n_{\text{iter}}$) and the regularization parameter $\lambda$ were first established by reconstructing two different undersampled datasets for each subject using various values for $n_{\text{iter}}$ and $\lambda$, and comparing the resulting reconstructions to the fully-sampled reference data. The set of parameter values ($\lambda = 0.07, n_{iter} = 20$) which consistently returned good reconstruction results was used for all other reconstructions: see Supporting Information Figure S1.

All reconstructions were performed off-line using an Intel (Intel, Santa Clara, CA, USA) Xeon CPU E5-2680 (v4) running at 2.40 GHz with 14 cores and 28 logical processors. A single-slab CS reconstruction took approximately 14 minutes.

### 2.4 Quantification of reconstruction quality

Previous work found that when comparing reconstructions from retrospectively undersampled MRA data to the corresponding fully-sampled data, the vessel-masked structural similarity (SSIM) index correlates very well with visual evaluation by radiologists [27]. The root-mean-square error, which is also often used in literature, was found to result in lower agreement [27] and is therefore not used in this work.

The structural similarity is a quantitative estimate of the perceived visual agreement between a set of images based on local intensity variations. Higher structural similarity indicates greater agreement between a set of images, with a value of 1 indicating perfect agreement. In this work, the SSIM is calculated for maximum intensity projections (MIPs) normalized to the 99[th] percentile of the pixel-wise intensities. The mean SSIM is calculated over a vessel-masked region (Figure 1b), which reduces the sensitivity to variations in the background signal.

The SSIM requires spatial consistency of the information in the two images being compared. This limits its use when comparing fully-sampled reconstructions to those from prospectively undersampled data, since minor subject motion between consecutive scans significantly influences the SSIM. Therefore, a metric ("number of detected peaks") to estimate the change in the number of visible vessels in an image (similar to the method introduced by Meixner et al. [9]) was used to assess the quality of prospectively undersampled data. From each brain-masked MIP (normalized to the 99$^{th}$ percentile), 100 cross-sectional lines in the left-right direction were taken, covering the central 50% of the field-of-view in the anterior-posterior direction. Along these lines, peaks in the intensity profiles were detected using the `findpeaks` function in Matlab with a minimum peak prominence of 0.15.

To quantify the difference in visibility of small vessels in prospectively undersampled acquisitions, a comparable approach was used to estimate the visibility of the LSAs in coronal projections.

# 3. Results

## 3.1 Fully-sampled data

Figure 1 shows MIPs of the fully sampled reconstructions of the central 2 (out of 4) slabs for both subjects in Cohort 1, which were used for parameter optimization. Those slabs were used as the ground-truth reference for all combinations of parameters used to assess the different retrospectively undersampled reconstructions.

## 3.2 Undersampling Optimization

The mean vessel-masked SSIM resulting from the retrospective undersampling parameter optimization for six acceleration factors R (5, 7, 9, 11, 13, 15) for all four slabs (in Figure 1) are presented in Figure 2. Each of the values in Figure 2 is the mean value for two different undersampling masks (using the same undersampling parameters), such that each datapoint is the average of 8 comparisons: four different slabs, with two different undersampling masks each. See Supporting Information Figures S2-S5 for the results for each individual slab.

Especially for higher acceleration factors, Figure 2 indicates an optimal set of undersampling parameters when using 12×12 calibration lines ("*calib* = 12") with a polynomial order of approximately 2.0 to 2.4. For datasets reconstructed from the Cohort 1 data with a lower spatial resolution (0.5 mm and 0.6 mm isotropic), optimal undersampling parameters consisted of 12×12 calibration lines and a polynomial order of approximately 1.6 to 2.0 (data not shown).

Figure 3Figure 4 show example reconstructions from fully-sampled data and prospectively undersampled data (Cohort 2), using the identified undersampling parameters as well as larger calibration region sizes as previously used [9].

Figure 5 shows examples of the implementation of the "number of detected peaks"-metric, as well as the resulting differences between the various undersampling schemes for the six volunteers in Cohort 2. Results are shown for whole-brain axial MIPs and coronal MIPs of the lenticulostriate region.

At R = 7.2, the mean relative number of detected peaks when using 12×12 compared to 32×32 calibration lines increases by 1.5 ± 1.0 percentage points (from -4.8% to -3.3%) for axial MIPs and by 7.2 ± 4.2 percentage points (-24.2% to -17.0%) for the LSAs. At R = 15, the number of detected peaks increases by 7.7 ± 0.5 percentage points (-25.3% to -17.7%) for axial MIPs and by 14.7 ± 1.6 percentage points (-59.7% to -45.0%) for the LSAs.

The change in the number of detected peaks when using optimized undersampling parameters is statistically significant at R = 15, based on a two-sample t-test: *p < 0.001* for both whole-brain and LSA MIPs. At R = 7.2, statistical significance was not reached, but trend values of *p = 0.17* and *p = 0.11* were obtained for whole-brain and LSA MIPs, respectively.

## 4. Discussion

Although differences in the image quality when using various auto-calibration region sizes and polynomial orders are visible for all acceleration factors in Figure 2, the relative importance of using optimized parameters increases for higher acceleration factors. For all acceleration factors, the combination of 12×12 calibration lines with a polynomial order of 2-2.4 consistently yielded the best reconstruction accuracy. Simulation results at reduced spatial resolutions indicate the same optimal calibration region size, but with slightly lower polynomial orders (approximately 1.6-2.0 at (0.6mm)$^3$). For all resolutions, optimization of the calibration region size has a bigger influence on the image quality than optimization of the polynomial order. This optimal calibration region size is considerably smaller than values found in literature. Reducing the calibration region size from 32×32 to 12×12 lines in k-space corresponds to an 86% reduction in the amount of scan time required for scanning this central k-space region. This makes it possible to spend more scan time acquiring data at higher spatial frequencies, improving the visibility of sparse image structures such as small vessels. Asymmetric calibration regions, with reduced coverage in the partition-direction ($k_z$), may allow further reductions but were not included in this work because of limitations in the used implementation of ESPIRiT.

Figure 3 and Figure **4** show a clear reduction in the number of visible vessels for both optimized and literature-based acquisition parameters at high acceleration factors when compared to a fully-sampled acquisition. However, the vessel visibility and sharpness noticeably improves when using optimized undersampling parameters versus literature parameters, especially at high acceleration factors. Although accurate quantitative comparison of the different prospectively undersampled acquisitions using SSIM is not possible due to subject motion between scans, the reduction in signal loss when using optimized parameters was quantitatively approximated using the detection of vascular signal peaks. The identified peak locations in reconstructions from prospectively undersampled data are consistent with the locations of the peaks in the corresponding fully-sampled data (not shown), indicating that the metric gives a representative approximation of relative image quality.

Prospectively and retrospectively undersampled data can differ in image quality because of the possibility of eddy-current artifacts due to larger gradient switching in prospectively undersampled acquisitions and different amounts of total subject motion due to the different scan times. However, the observed improvement in image quality when using optimized undersampling schemes identified using the retrospectively undersampled data (Figure 2) is consistent with the results found in Figure 3-Figure **5** for prospectively undersampled data. This applies to both the improvement in image quality when using smaller calibration regions (in particular, 12×12 calibration lines) and to the increased importance of acquisition parameter optimization at higher acceleration factors. This improvement when using optimized undersampling parameters can be achieved without increasing scan time or reconstruction time, and without additional technical requirements.

For one of the six volunteers in cohort 2, an increase in the number of detected peaks is visible in the undersampled acquisitions (purple data in Figure 5c and Figure **5**e). This is likely caused by a reduction in image quality in the fully-sampled dataset due to subject motion during that scan, highlighting a benefit of accelerated acquisitions.

The data used in this work were acquired using a relatively simple protocol, which does not make use of techniques such as additional (e.g. fat or venous) signal saturation [9], [28-29], parallel transmission [30-31], VERSE-shimming [9], [28], [32], or intravenous contrast agents [33-34]. Such techniques can enhance the contrast and thereby improve the CS reconstruction results. Alternative protocols may result in changes to the acquired contrasts. However, the optimal undersampling parameters were found to be consistent for volumes with high differences in vascular characteristics and visibility (Figure 2), and are expected to remain consistent for different contrasts. The improvement of image quality when using 12×12 calibration lines was also found to be consistent for data acquired using various acceleration factors and spatial resolutions. Although 32-channel receive coils are most commonly used in 7T MRI, it remains unclear how the results presented here would translate to different coil configurations.

Previous work on the optimization of acquisition parameters for CS T1-weighted MRI in 3D found that optimized sampling schemes require increasingly dense sampling in the centre of k-space for higher acceleration factors [19], and that the extent of the calibration region should be as high as possible for 2D-MRI [17]. This is in contrast to the results found here, with an optimized set of acquisition parameters which appears to be consistent for all acceleration factors and which contains a small calibration region. This difference may be explained by the inherently sparser image contrast of MRA compared to T1-weighted MRI, and the use of three-dimensional instead of two-dimensional k-space data. Since this sparser image signal is contained in high-frequency areas of k-space, sampling at off-centre locations of k-space remains important for higher acceleration factors in the case of MRA.

# 5. Conclusion

Optimized undersampling parameters for 3D MRA at 7T using compressed sensing reconstruction were established. For all acceleration factors, the highest image quality was achieved by using a fully-sampled calibration area of 12×12 lines and a polynomial order of around 2. Although the optimized undersampling parameters were the same for all acceleration factors, the importance of using optimized undersampling parameters was found to increase for higher acceleration factors.

**Acknowledgements**

The Wellcome Centre for Integrative Neuroimaging is supported by core funding from the Wellcome Trust (203139/Z/16/Z). We also thank the Dunhill Medical Trust and the NIHR Oxford Biomedical Research Centre for support (PJ). MdB acknowledges studentship support from Siemens Healthineers and the Dunhill Medical Trust. AH acknowledges support from the BHF Centre of Research Excellence, Oxford (RE/13/1/30181).

**Conflict of Interest**

P.J. is the Editor-in-Chief of Magnetic Resonance in Medicine. In line with COPE guidelines, he recused himself from all involvement in the review process of this paper, which was handled by an Associate Editor. He and the other authors have no access to the identity of the reviewers.

# Figures

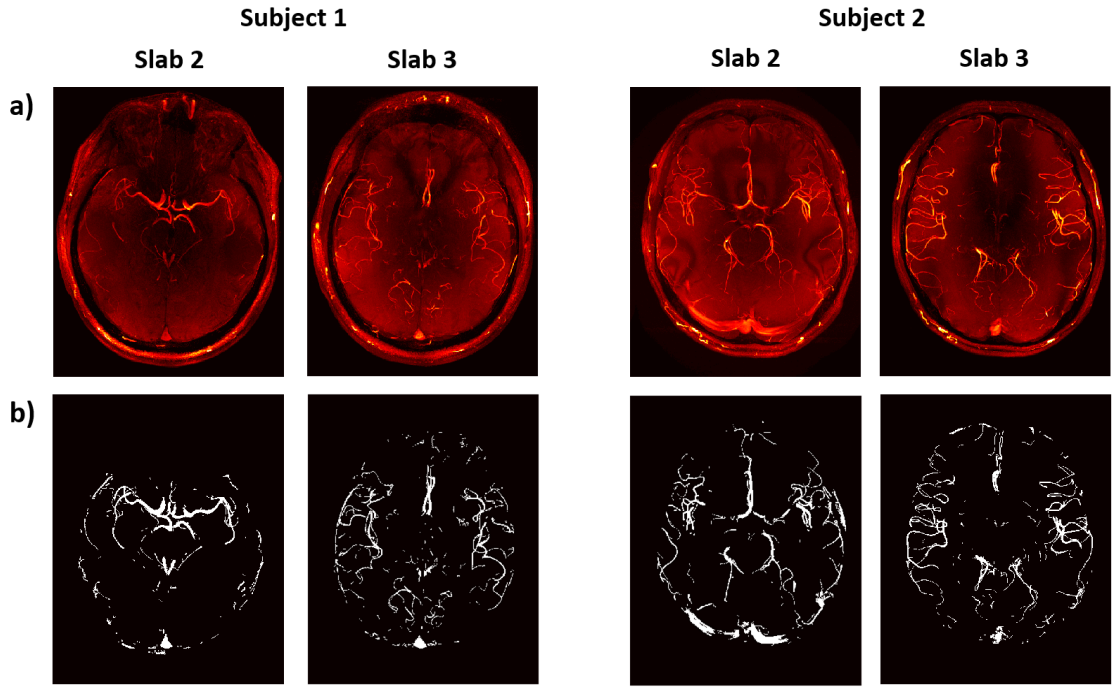

*Figure 1:* The four slabs used to test retrospective undersampling strategies (from Cohort 1). *(a)* MIPs of the fully-sampled reconstruction for each slab. *(b)* The corresponding vessel masks, used for computing the vessel-masked SSIM [27].

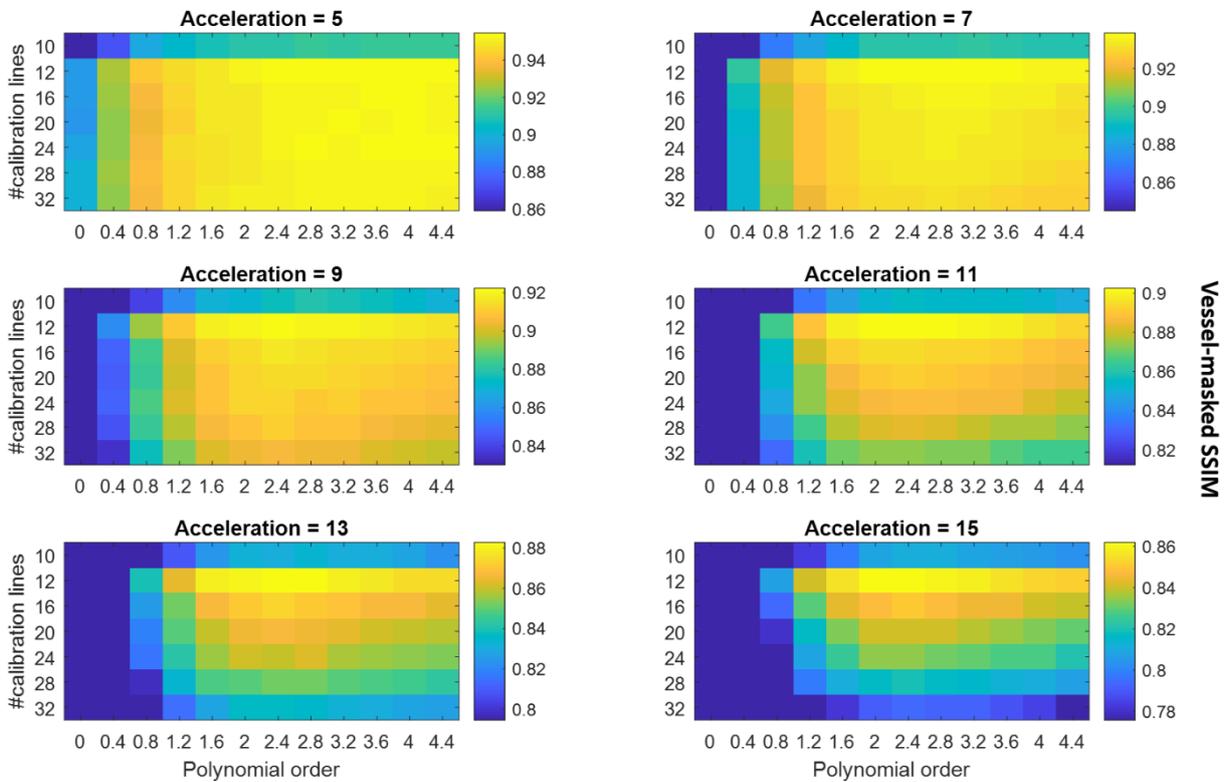

*Figure 2:* Average SSIM values for various sets of undersampling parameters. Each datapoint represents the mean vessel-masked SSIM values of the four slabs shown in Figure 1. The results for each of the four slabs are separately shown in the Supporting Information. The

*scaling of the SSIM-values in each individual figure runs from the maximum value (for the given acceleration factor) to 90% of that maximum value, to maximize the visibility of the relative image quality for each acceleration factor.*

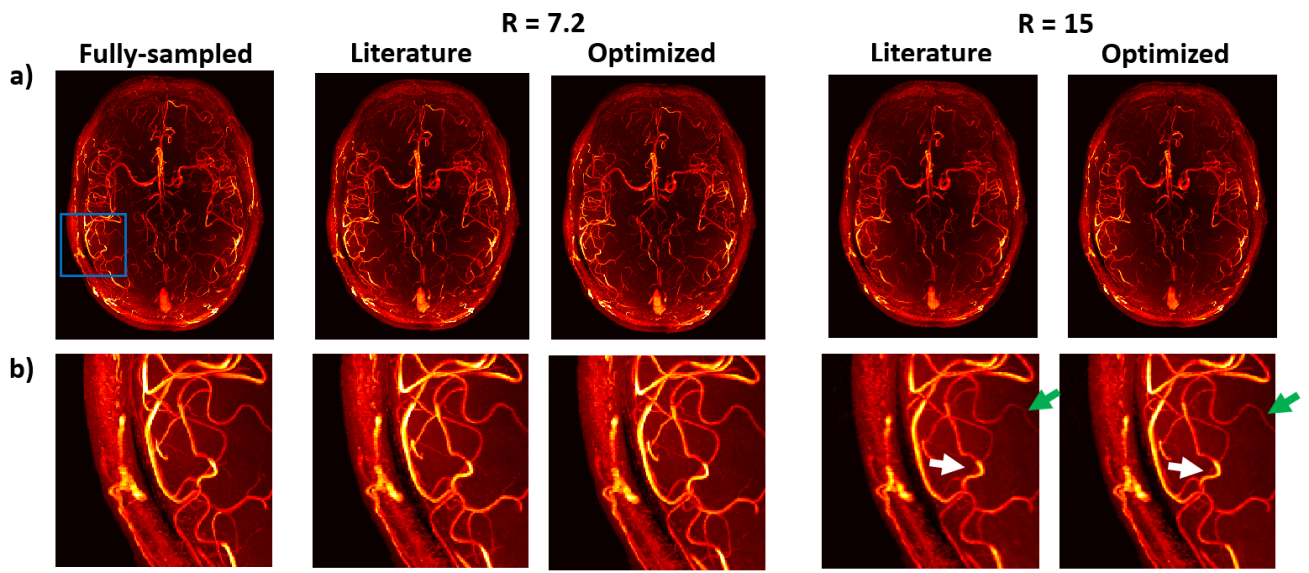

***Figure 3:*** *Comparison of axial MIPs from optimized (calib = 12) and literature-based (calib = 32) undersampling schemes for R = 7.2 and R = 15.* ***(a)*** *Reconstructed images from fully-sampled data and the different prospectively undersampled acquisitions.* ***(b)*** *Close-up of the region marked with a blue square in (a), for all acquisitions. Green arrows indicate examples of improved vessel visibility when using optimized undersampling parameters; white arrows indicate improved vessel sharpness.*

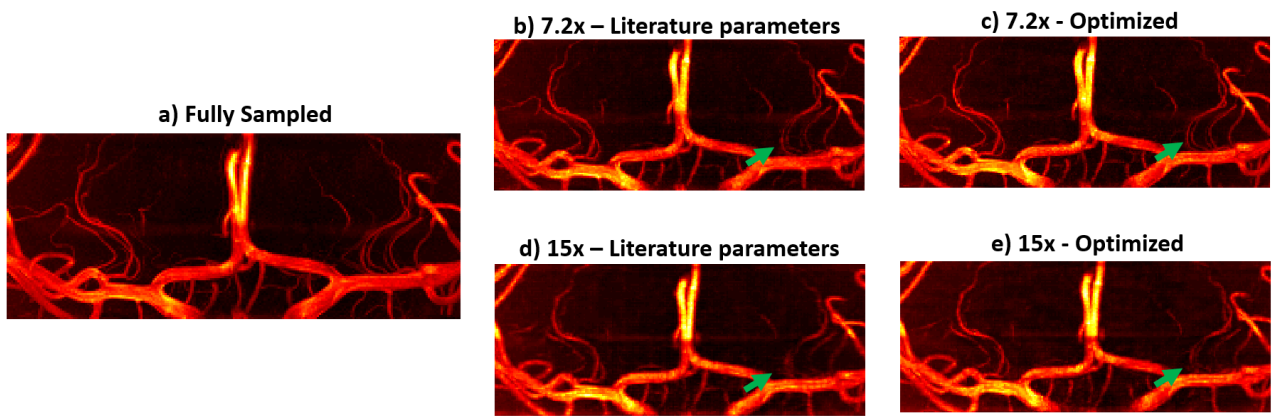

***Figure 4:*** *Comparison of coronal MIPs of the LSAs from optimized and literature-based undersampling schemes. Images shown for* ***(a)*** *fully-sampled data;* ***(b-c)*** *data for R = 7.2 using literature-based (b) and optimized (c) prospectively undersampled acquisitions; and* ***(d-e)*** *data for R = 15 using literature-based (d) and optimized (e) prospectively undersampled acquisitions.*

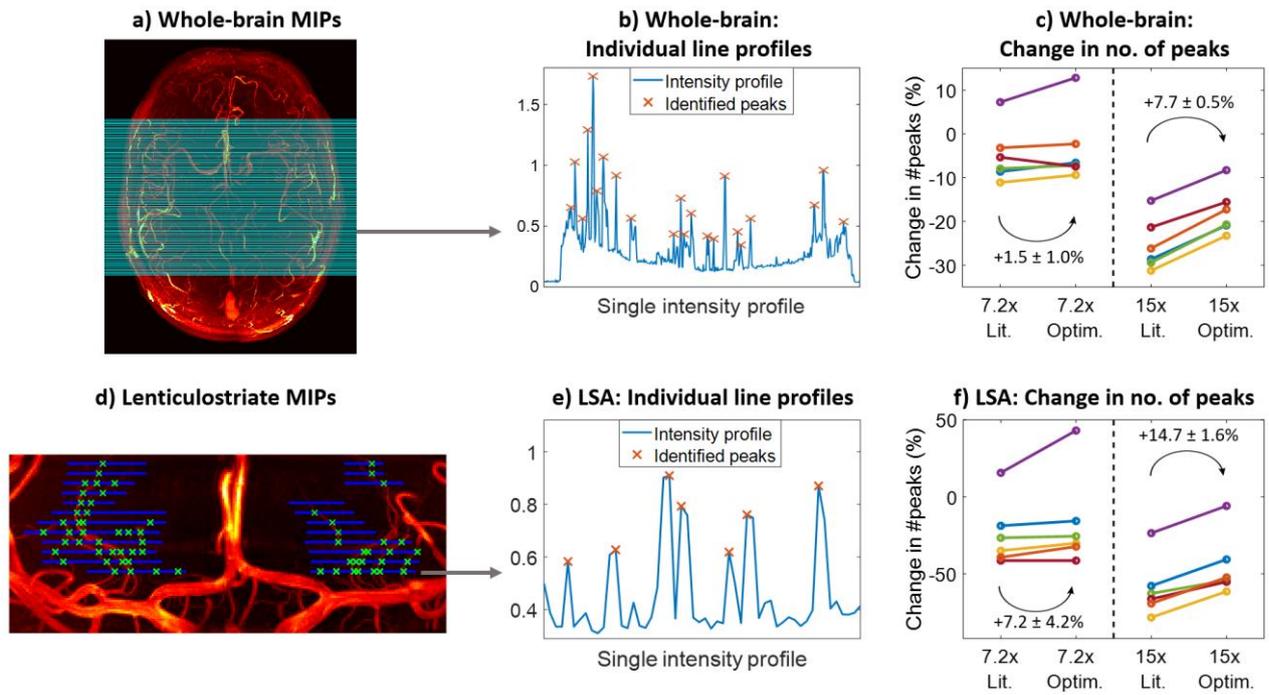

*Figure 5:* *Quantification of the change in the number of detected peaks. **(a-c):** Whole-brain MIPs. **a)** The 100 lines along which peaks in the intensity profiles were identified on all datasets (as shown in **(b)** for Line 25). **(c)** The change in the identified number of peaks for optimized ("Optim.") and literature-based ("Lit.") undersampled acquisitions, relative to fully-sampled acquisitions. Each colour indicates a subject from Cohort 2. **(d-f):** The same as (a-c) for MIPs of lenticulostriate ROIs. Line locations were drawn manually at 5-pixel intervals.*

# Supporting Information

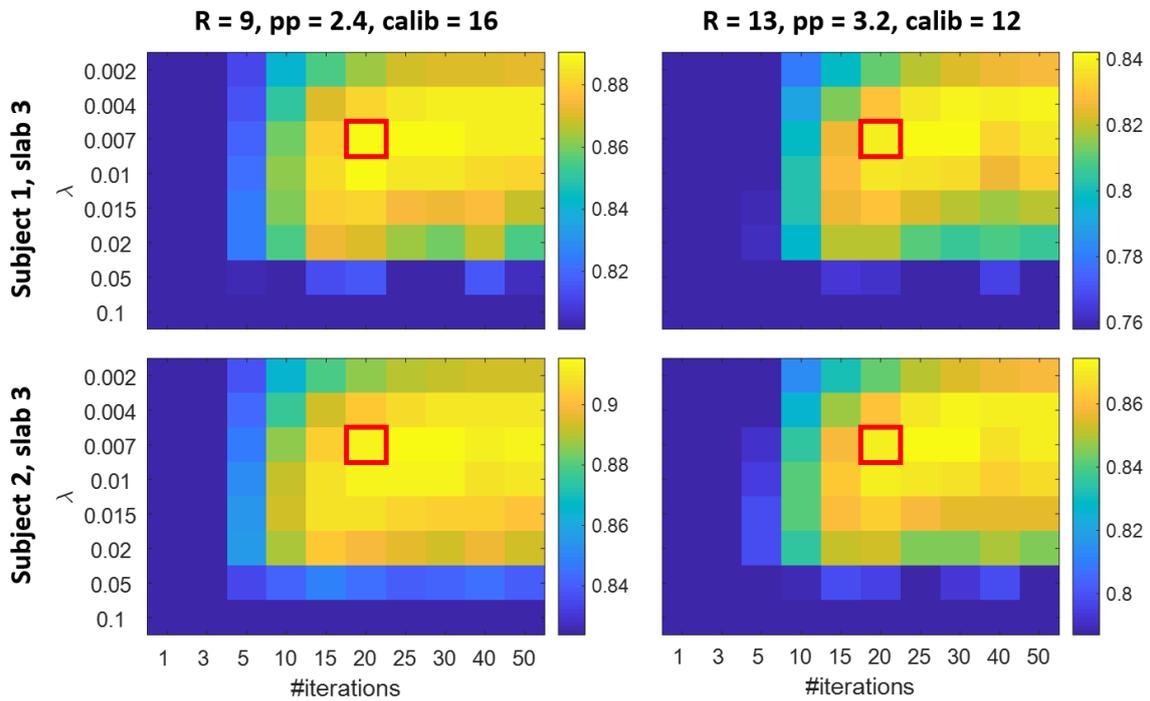

***Supporting Information Figure S1***: Results of the initial reconstruction parameter optimization, used to estimate appropriate values for $n_{iter}$ and the regularization parameter $\lambda$. Reconstruction accuracy is given using the mean vessel-masked SSIM. Results are shown for two different imaging volumes, each undersampled using two different sets of undersampling parameters. Red boxes indicate the set of reconstruction parameters ($\lambda =$ 0.07, 20 iterations) used for all later reconstructions because of the consistently good results for all cases, within reasonable reconstruction times.

# Subject 1, slab 2

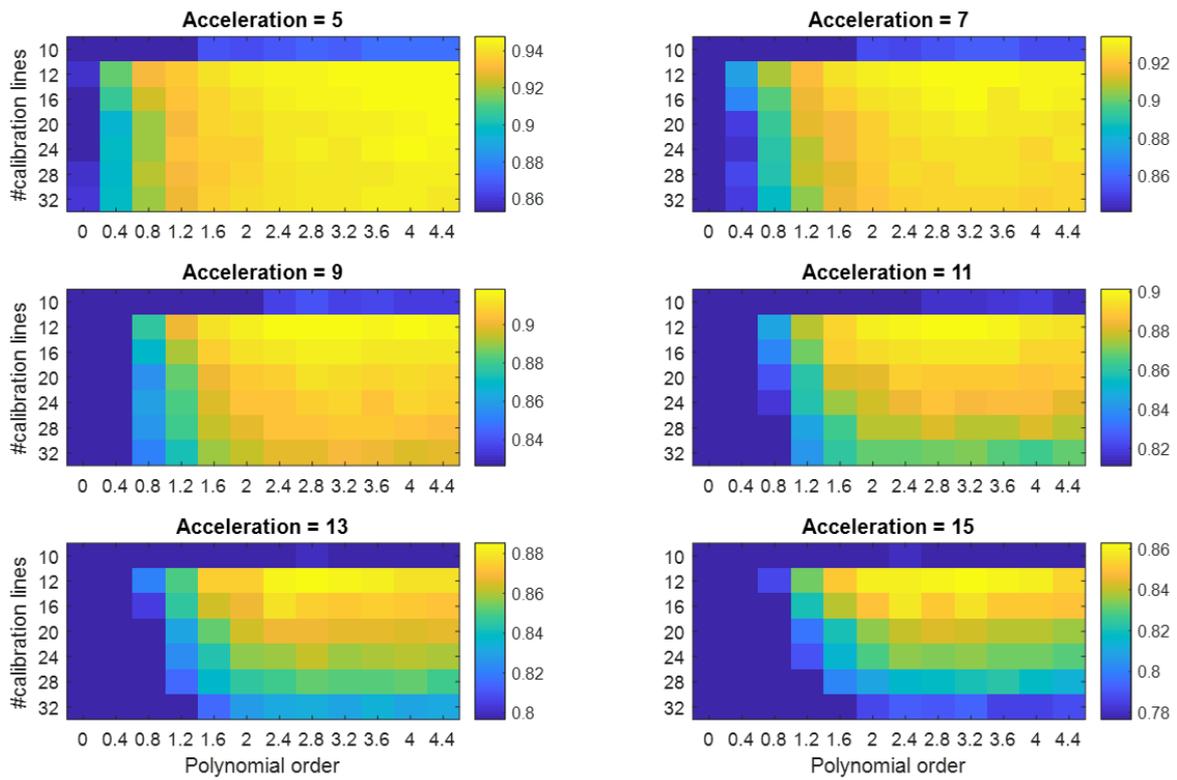

***Supporting Information Figure S2:*** *The vessel-masked SSIM-values comparing the fully sampled reference datasets and corresponding reconstructed datasets for Subject 1, Slab 2.*

# Subject 1, slab 3

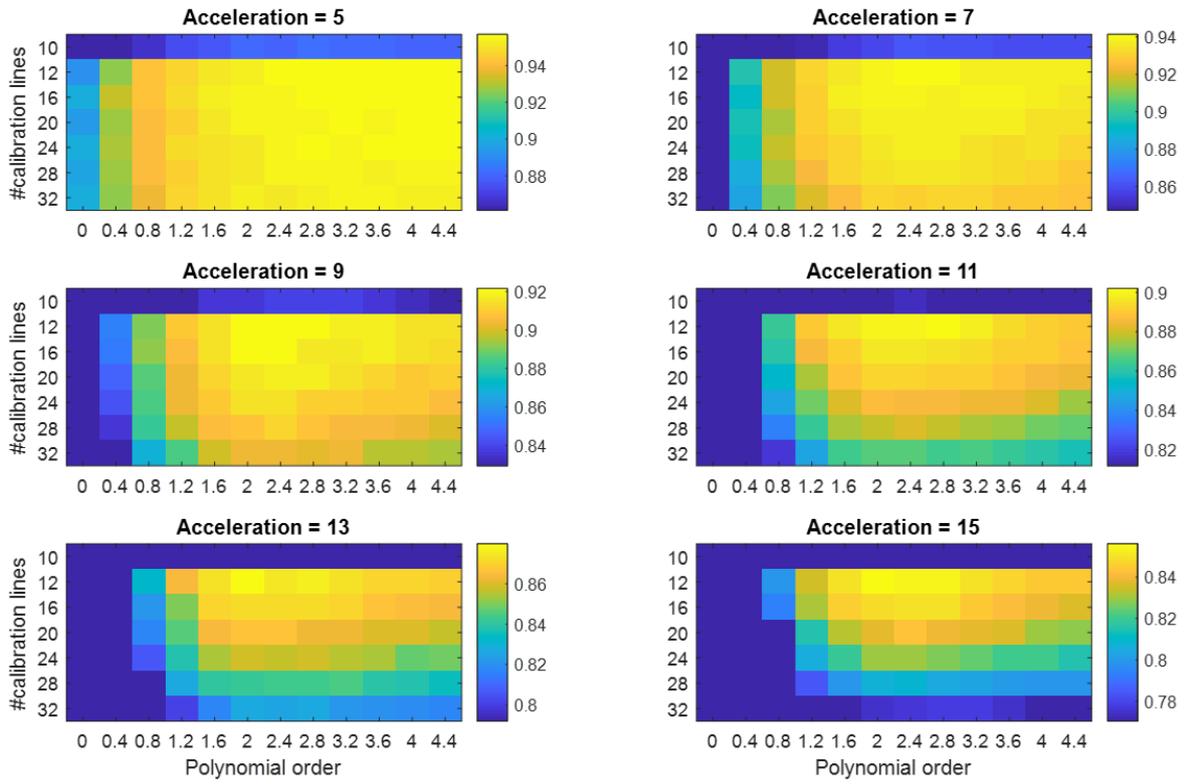

***Supporting Information Figure S3:*** *The vessel-masked SSIM-values comparing the fully sampled reference datasets and corresponding reconstructed datasets for Subject 1, Slab 3.*

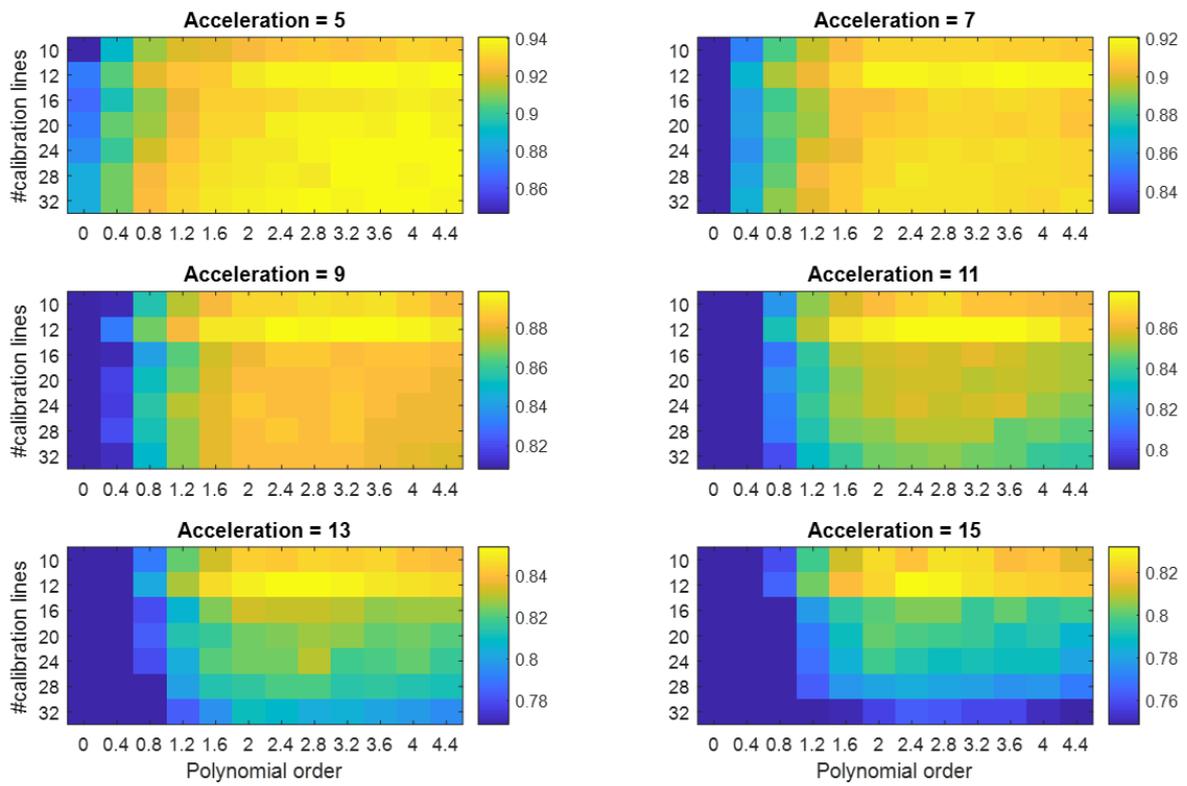

*Supporting Information Figure S4*: The vessel-masked SSIM-values comparing the fully sampled reference datasets and corresponding reconstructed datasets for Subject 2, Slab 2.

# Subject 2, slab 3

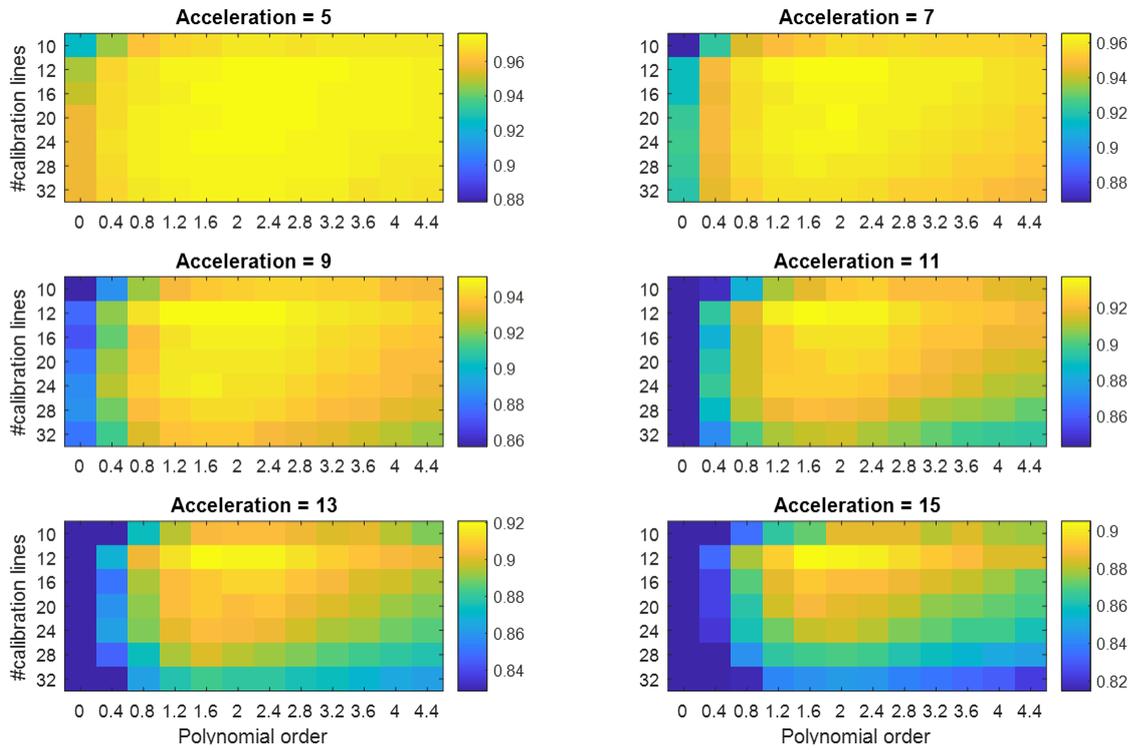

***Supporting Information Figure S5***: *The vessel-masked SSIM-values comparing the fully sampled reference datasets and corresponding reconstructed datasets for Subject 2, Slab 3.*